\documentclass[namedreferences]{SolarPhysics}
\usepackage[optionalrh]{spr-sola-addons} 
\usepackage{graphicx}                    
\usepackage{amssymb}                    
\usepackage{color}                       
\usepackage{url}                         

\newcommand{\aap}{    {\it Astron. Astrophys.}}

\newcommand{\apj}{    {\it Astrophys. J.}}
\newcommand{\apjl}{    {\it Astrophys. J. Lett.}}

\newcommand{\grl}{    {\it Geophys. Res. Lett.}}

\newcommand{\jgr}{    {\it J. Geophys. Res.}}

\newcommand{\solphys}{{\it Solar Phys.}}

\newcommand{\ssr}{    {\it Space Sci. Rev.}}

\newcommand{\degree}{$^{\circ}$~}
\newcommand{\degreee}{$^{\circ}$}

\newcommand{\kmsec}{km~s$^{-1}$}

\begin{document}

\begin{article}

\begin{opening}

\title{Speeds and arrival times of solar transients approximated by self-similar expanding circular fronts}

%
\author{C.~\surname{M\"ostl}$^{1,2}$ and 
        J.A.~\surname{Davies}$^{3}$}

%
\runningauthor{M\"ostl and Davies}
\runningtitle{Arrival time calculation for self-similar expanding transients}

%
  \institute{$^{1}$ Space Science Laboratory, University of California, Berkeley, CA 94720, USA.\\‚
             $^{2}$ Institute of Physics, University of Graz, A-8010, Austria.\\email: \url{christian.moestl@uni-graz.at}\\
          $^{3}$ RAL Space, Harwell Oxford, Didcot, OX11 0QX, UK.
              }

\begin{abstract}

The NASA Solar TErrestrial RElations Observatory (STEREO) mission opened up the possibility to forecast the arrival times, speeds and directions of solar transients from outside the Sun-Earth line. In particular, we are interested in predicting potentially geo-effective Interplanetary Coronal Mass Ejections (ICMEs) from observations of density structures at large observation angles from the Sun (with the STEREO Heliospheric Imager instrument). We contribute to this endeavor by deriving analytical formulas concerning a geometric correction for the ICME speed and arrival time for the technique introduced by Davies \emph{et al.} (2012, \emph{Astrophys. J.}, in press) called Self-Similar Expansion Fitting (SSEF). This model assumes that a circle propagates outward, along a plane specified by a position angle (e.g. the ecliptic), with constant angular half width ($\lambda$).  This is an extension to earlier, more simple models: Fixed-$\Phi$ Fitting ($\lambda = 0$\degreee) and Harmonic Mean Fitting ($\lambda=90$\degreee). This approach has the advantage that it is possible to assess clearly, in contrast to previous models, if a particular location in the heliosphere, such as a planet or spacecraft,  might be expected to be hit by the ICME front. Our correction formulas are especially significant for glancing hits, where small differences in the direction greatly influence the expected speeds (up to $100-200$~\kmsec) and arrival times (up to two days later than the apex). For very wide ICMEs ($2\lambda>120$\degreee), the geometric correction becomes very similar to the one derived by M\"ostl \emph{et al.} (2011, \emph{Astrophys. J.} 741, id. 34) for the Harmonic Mean model. These analytic expressions can also be used for empirical or analytical models to predict the 1 AU arrival time of an ICME by correcting for effects of hits by the flank rather than the apex, if the width and direction of the ICME in a plane are known and a circular geometry of the ICME front is assumed.
\end{abstract}

%

\end{opening}


%
 \section{Introduction}

Predicting the effects of solar eruptions at Earth is a scientific art in the making. Many different approaches have been developed world-wide, using analytical, numerical or empirical techniques, to better understand the physics behind Interplanetary Coronal Mass Ejections (ICMEs) which should ultimately lead to space weather predictions on a routine basis with reasonable accuracy. The NASA Solar TErrestrial RElations Observatory (STEREO, \citeauthor{kai08} \citeyear{kai08}), launched in 2006 and consisting of two observatories leading and lagging the orbit of Earth by 22.5\degree per year, plays a major role in these efforts. STEREO has been designed to enhance our ability to forecast ICME arrival times at Earth through more precise determination of ICME speeds and directions with both stereoscopic and single-spacecraft techniques, based on observations taken from outside the Sun-Earth line to get rid of projection effects \cite{gop01,sch05} and filling the imaging gap between the Sun and 1 AU with the Heliospheric Imagers (HI) instruments \cite{eyl09}. The latter has also been achieved from within Earth orbit by the Solar Mass Ejection Imager (SMEI) instrument \cite{eyl03}.

An idea originally introduced by \citeauthor{she99} (\citeyear{she99})
 was used to predict directions and speeds of ICMEs \cite{dav09} and co-rotating interaction regions (\citeauthor{rou08} \citeyear{rou08}) from HI images. This method makes use of a simple geometrical fact, which can be illustrated as follows: if someone throws a ball with constant velocity along a constant direction, an observer measuring the angle between himself, the thrower and the ball as a function of time will notice deceptive decelerations or accelerations in the time-angle profile. The form of this function depends on the speed and direction of the ball with respect to the observer. Applied to ICMEs in any plane given by a position angle to solar north (e.g. the ecliptic plane), these profiles, known analytically for either point-like (Fixed-$\Phi$, FP,  \citeauthor{she99} \citeyear{she99}; \opencite{kah07}) or circular fronts (Harmonic Mean, HM, \opencite{lug09a}; \opencite{how09a}), can be inversely fitted through a minimization process to observed profiles of the elongation angle of the ICME front from the Sun (e.g. \citeauthor{rou09} \citeyear{rou09}; \citeauthor{dav09} \citeyear{dav09};  \citeauthor{how09a}, \citeyear{how09a}; \citeauthor{tap09}, \citeyear{tap09};
 \citeauthor{dav09b} \citeyear{dav09b}; \citeauthor{sav09} \citeyear{sav09,sav10}; \citeauthor{moe09c} \citeyear{moe09c,moe10,moe11}; \citeauthor{lug10b}, \citeyear{lug10b}). We denote the techniques as FPF and HMF. The FPF technique has been successfully applied in real time to an ICME in April 2010 \cite{dav11}, and to some other recent real time events, and is a candidate for a technique to be used routinely on a future space weather mission at the L5 point in the Sun-Earth system \cite{gop11} or on \emph{Solar Orbiter}. In comparison to a numerical simulation, some biases have been found for FPF/HMF, mainly for cases where the ICME is either close to the limb or heading towards the observer \cite{lug11}. Similarly to the above mentioned techniques, stereoscopic versions exist for FP (\citeauthor{liu10} \citeyear{liu10}a) and HM \cite{lug10}, relaxing the assumptions of constant speed and direction, though these techniques have not yet revealed strong deviations from radial propagation far from the Sun. 

An improvement to the single-spacecraft techniques has been introduced by  (\citeauthor{dav12} \citeyear{dav12}, further called DA12), called self-similar expansion or SSE to convert elongation to distance. This is based on model 2 by 
\citeauthor{lug10} (\citeyear{lug10}), and DA12 made it suitable as a prediction tool through inverse fitting (SSEF). This model has a great appeal by solving some problems (see discussion by \citeauthor{moe11} \citeyear{moe11}, further: MO11) which are intrinsic to the point-like or very wide front assumptions of FP and HM, respectively, especially when thinking about when and with which speed the ICME front will hit a planet, resulting in possible magnetic storms, or a spacecraft which makes \emph{in situ} observations of the solar wind plasma and magnetic field parameters. 

The aim of this short report is to clarify the calculation of arrival times and speeds at a particular position in the heliosphere with the self-similar expansion model (DA12) or its stereoscopic version, model 2 in \citeauthor{lug10} (\citeyear{lug10}). The assumption behind it is that the ICME front can be approximated by a circle with constant angular width (see Figure~\ref{deriv}). Our term ``front'' leaves open whether one wants to investigate the speed and arrival time of the interplanetary shock wave or the front boundary of the magnetic flux rope or ejecta driving the shock. Given the position of a spacecraft with respect to the direction of the ICME apex (the point along the front with greatest heliospheric distance from the Sun), this model will lead to later arrival times and lower speeds, or even no hit at all. In this way, we extend the work of MO11 who derived a similar arrival time correction for the HM model, which is a limiting case of SSE for large ICME width. The FP model can also easily be understood as the limiting case of SSE for negligible width. Clearly, in the real solar wind environment it is questionable if these idealized conditions are even roughly met, and distortions of the fronts will be likely. Also note that a spherical expanding front centered on the Sun (called ``Point-P'' by various authors) has the same heliocentric distance‚ everywhere, and corrections to speed and arrival time are not needed.  However, to derive geometrically consistent arrival times and speeds with SSE or its triangulation version it is necessary to use the formulas derived in this paper.

The analytical considerations we present in this paper are not restricted to techniques used on HI instruments, but they provide a simple framework to assess the effects of an ICME hitting a target with its apex or flank. If the direction of an ICME is known from stereoscopic observations close to the Sun from various techniques \cite{the09,tap09,tem09,dek09,mie10,byr10,rod11}, an educated  guess can be made regarding the difference in heliospheric longitude between the ICME apex and the spacecraft or planet. A model for ICME propagation can then be used to extrapolate the arrival time and speed of the ICME apex to 1 AU (e.g. \citeauthor{gop01} \citeyear{gop01}; \citeauthor{vrs02}, \citeyear{vrs02}; \citeauthor{sch05} \citeyear{sch05}; \citeauthor{sis06}, \citeyear{sis06}; \citeauthor{vrs10} \citeyear{vrs10}; \citeauthor{mal10}, \citeyear{mal10}; \citeauthor{vrs12} \citeyear{vrs12}, this issue), which can subsequently be corrected for the longitude difference using the formulas presented here, given that the width of the ICME is known, or simply assumed.

The average width of a CME in coronagraph observations is known to be about 50\degree \cite{yas04} with considerable scatter around this value. Direct comparison of STEREO/HI movies to a rendered 3D density model of a loop-like ICME revealed a true (de-projected) width of 70\degree \cite{woo10} for one event. However, we are interested here in the angular extent of the ICME front along the line-of-sight (LOS) in a given plane (such as the ecliptic). Stereoscopic modeling assuming such a geometry by 
\citeauthor{lug10} (\citeyear{lug10}) revealed half (full) widths in the ecliptic from $25-45$\degree ($50-90$\degreee), and observations of interplanetary shocks and their solar sources point to extents of up to 100\degree \cite{ric93}. One should also be aware that the ICME angular extent along the LOS should depend upon its orientation with respect to the ecliptic plane (e.g. \citeauthor{liu10b} \citeyear{liu10b}b; \citeauthor{kil11} \citeyear{kil11}; \citeauthor{moe11} \citeyear{moe11}).

We start by showing the derivation of the formulas in Section 2 and discuss them in comparison to the former FP and HM models. We then proceed to plot the effects on arrival time and speed predictions in the next two sections, discuss their significance and conclude with comments on our assumptions and future work.


\section{Derivation of arrival time and speed for SSE}

\begin{figure}[t] 
\centering
    \includegraphics[width=0.9\textwidth,clip=]{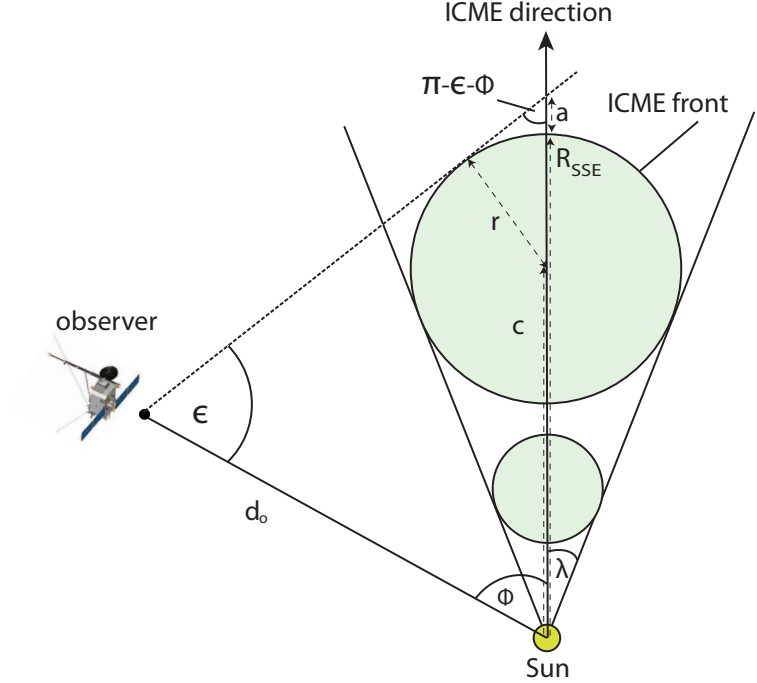}
    \caption{The geometry of the SSE model. An observer looks along the tangent to a circular front at an elongation $\epsilon(t)$ to Sun center. The circle expands  with constant angular half-width $\lambda$ along a constant ICME propagation direction $\phi$. Note that the full width of the ICME is $2 \lambda$. }\label{deriv}
\end{figure}

Figure~\ref{deriv} shows the geometry introduced by \inlinecite{lug10} and elaborated by DA12. It is assumed that an observer looks along the tangent to a circular front propagating radially away from the Sun with constant angular width. The elongation-to-distance conversion for SSE can be derived as: 
\begin{eqnarray}
R_{SSE}=\frac{d_o \sin({\epsilon(t)})(1+\sin\lambda)}{\sin(\epsilon(t)+\phi) +\sin\lambda}
\label{sse}
\end{eqnarray}
with $d_o$ the radial distance of the observing spacecraft (called $O$) from the Sun, $\epsilon(t)$ the measured elongation angle as function of time,  $\lambda$ the angular half-width of the ICME, angle $\phi$ the propagation direction as measured from the observer and $R_{SSE}$ the heliocentric distance of the ICME apex. For the large and small triangles in Figure~\ref{deriv} we can write: 
\begin{eqnarray}
\frac{\sin \epsilon(t)}{R_{SSE}+a}=\frac{\sin (\pi-\epsilon-\phi)}{d_o}, \label{eq:sse1}\\
\sin(\pi-\epsilon-\phi)=\frac{r }{r+a}.\label{eq:sse2}
\end{eqnarray}
We further need a relation between $R_{SSE}$ and $r$, the radius of the circle. Using $R_{SSE}=r+c$ and $r=c \sin \lambda$ leads to: 
\begin{eqnarray}\label{eq:curvature}
r (R_{SSE}, \lambda)= \frac{R_{SSE} \sin \lambda}{(1+ \sin \lambda)}, 
\end{eqnarray}
and can be understood as a radius of curvature, which is, however, coupled to the width $\lambda$. It also  is a function of the apex position, and thus time, because the circle expands as the front propagates outwards. Using Equation~\ref{eq:curvature} in Equation~\ref{eq:sse2} and solving for $a$, allows one to put $a$ into Equation~\ref{eq:sse1}, and after some rearranging, Equation~\ref{sse} is obtained. For completeness, the circle's central heliocentric distance is:
\begin{eqnarray}
c=\frac{R_{SSE}(t)}{(1+\sin{\lambda})}.
\end{eqnarray}
By comparison, the HM model assumes a circle which is attached to the Sun at all times \cite{lug09a,how09a}, and this simply corresponds to Equation~\ref{sse} for the case of $\lambda=90$\degreee:
\begin{eqnarray}
R_{HM}=\frac{2 d_o \sin{\epsilon(t)}}{\sin(\epsilon(t)+\phi) +1}
\label{hm}
\end{eqnarray}
If the width is negligible ($\lambda=0$\degreee), the FP conversion for elongation-to-distance follows from Equation~\ref{sse} (\citeauthor{she99} \citeyear{she99}):
\begin{eqnarray}
R_{FP}=\frac{d_o \sin{\epsilon(t)}}{\sin(\epsilon(t)+\phi)}
\label{fp}
\end{eqnarray}

\begin{figure}[t] 
\centering
    \includegraphics[width=0.9\textwidth,clip=]{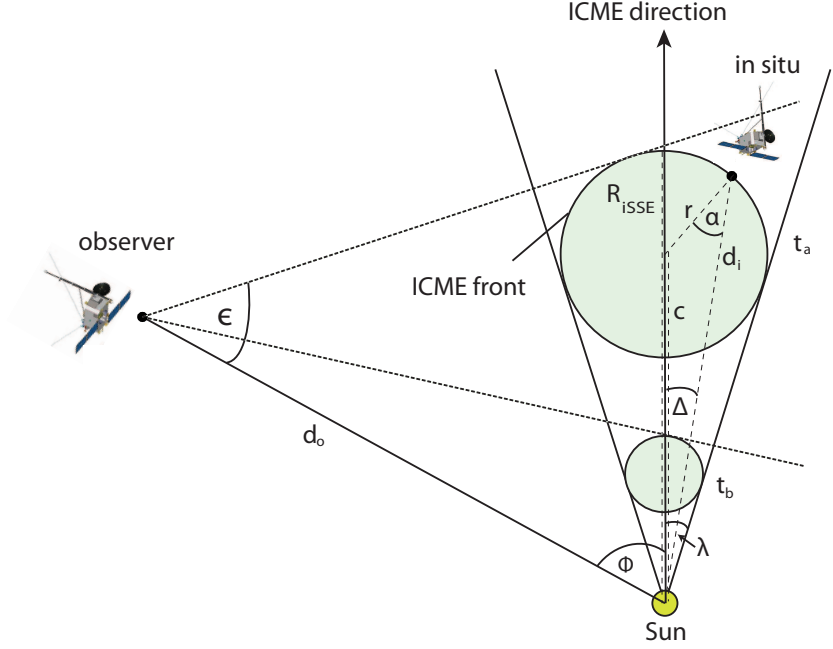}
    \caption{Similar to Figure 1, but now with an \emph{in situ} spacecraft $I$ situated at a heliocentric distance $d_i$ and at an angle $\Delta$ with respect to the ICME apex .}\label{sketch}
\end{figure}

\begin{figure}[ht] 
\centering
    \includegraphics[width=0.8\textwidth,clip=]{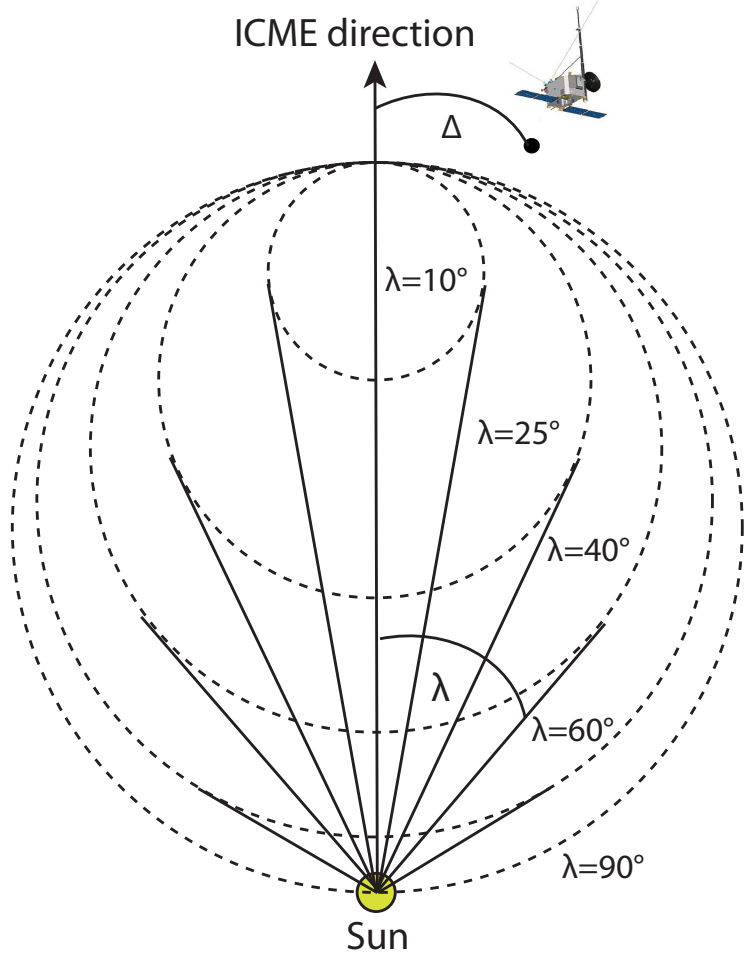}
    \caption{Illustration of five SSE fronts with different angular half widths $\lambda$. The \emph{in situ} observing spacecraft is positioned at an angle $\Delta$ to the ICME apex.}\label{size}
\end{figure}

We now ask ourselves: how far does the apex of the ICME front need to travel away from the Sun until it hits the \emph{in situ} spacecraft or planet $I$? Clearly, for a circular front this will be always greater than $I$'s heliocentric distance, $d_i$. As seen from Figure~\ref{sketch}, the ICME apex is separated at an angle $\Delta$ to $I$, so we make the following ansatz:
\begin{eqnarray}
\frac{\sin \Delta}{r}=\frac{\sin \alpha}{c},\label{eq:ansatz1}\\
\frac{\sin \Delta}{r}=\frac{\sin (\pi - \alpha-\Delta) }{d_i}.\label{eq:ansatz}
\end{eqnarray}
Using $r=c \sin \lambda $ in Equation~\ref{eq:ansatz1} results in:  
\begin{eqnarray}\label{eq:alpha}
\alpha = \arcsin \left (\frac{\sin \Delta}{ \sin \lambda} \right).
\end{eqnarray}
Substituting Equations~\ref{eq:alpha}  and~\ref{eq:curvature} into Equation~\ref{eq:ansatz} leads to: 
\begin{eqnarray}
R_{iSSE}(\Delta, \lambda, d_i)=\frac{d_i \sin \Delta (1+ \sin\lambda)}{\sin(\arcsin (\sin\Delta/ \sin\lambda)+\Delta)\sin\lambda}
\end{eqnarray}
Using trigonometric identities, this can be simplified to:
\begin{eqnarray}\label{eq:risse}
 R_{iSSE}(\Delta, \lambda, d_i) = \frac{d_i (1+\sin\lambda)}{\cos \Delta +
 \sqrt{\sin^2 \lambda - \sin^2 \Delta}}
\end{eqnarray}
We designated this special distance as $R_{iSSE}$, because it is the heliocentric distance of the ICME apex at the arrival time $t_a$ when the spacecraft is hit by the front, as illustrated in Figure~\ref{sketch}.
Again, for HM with $\lambda=90$\degreee,  Equation~\ref{eq:risse} indeed reduces to: 
\begin{eqnarray}
R_{iHM}=\frac{d_i}{\cos\Delta},\label{eq:rihm}
\end{eqnarray}
similar to the one used by MO11. For FP, one may assume that a small spherical front is centered around the assumed point, so simply $R_{iFP}=d_i$.

\subsection{Arrival time}

 If we further assume that the ICME front travels with constant speed $V_{SSE}$ for all times when $R_{SSE}(t) > d_i$,  we can express the arrival time correction $t_{cSSE}>0$ as:
\begin{eqnarray}\label{eq:tisse}
t_{cSSE}=\frac{R_{iSSE}(\Delta, \lambda, d_i)-d_i}{V_{SSE}}.
\end{eqnarray}
This formula states for how long the ICME apex has to travel, after it passed $d_i$, until $I$ is hit by the front. Here, $V_{SSE}$ can also be seen as an asymptotic speed given by an ICME propagation model.

For the SSE fitting technique (DA12), which results in constants for $\lambda$, $\phi$,  $V_{SSE}$, and the launch time $t_{0SSE}$, the arrival time at the spacecraft $I$ can be expressed as:
\begin{eqnarray}\label{eq:tasse}
t_{aSSE}=t_{0SSE} +\frac{R_{iSSE}(\Delta, \lambda, d_i)}{V_{SSE}}.
\end{eqnarray}
It needs to be understood that the launch time $t_{0SSE}$ is not determined from chromospheric or low coronal observations, but that it is rather a back-projection to the center of the Sun (where $\epsilon(t)=0$) to obtain a time of reference to be able to calculate $t_{aSSE}$ (see also MO11). It is assumed here that the speed $V_{SSE}$ is a constant for the range of elongation angles for which the time-elongation track, which is fitted with SSEF, is extracted from HI observations. Any earlier accelerations or decelerations of the CME closer to the Sun do not affect the calculation of $t_{aSSE}$. However, the launch time $t_{0SSE}$ will consequently be only a rough estimate of the real launch time of the CME in the corona.

 Similarly, for the HM model the arrival time calculation can be written as: 
\begin{eqnarray}\label{eq:tahm}
t_{aHM}=t_{0HM} +\frac{R_{iHM}(\Delta,  d_i)}{V_{HM}}.
\end{eqnarray}
which reduces with Equation~\ref{eq:rihm} to an expression consistent with MO11 (their Equation~B7).

Note that for SSE, not only is the arrival time well defined, but the prediction if a spacecraft or planet is hit by an ICME or not is given by the conditions $\lambda > \Delta$ (hit) and $\lambda < \Delta$ (no hit). This is much more precise as compared to FP, where, strictly, a point never hits another point. Also, for HM, which corresponds to SSE for $\lambda=90^{\circ}$, the circle will always hit if the \emph{in situ} spacecraft is inside the half-space given by $\phi \pm 90^{\circ}$.


\subsection{Arrival speed}

Additionally, for an expanding circular front, every point along the front moves with a slower speed compared to the apex. Thus a prediction of a corrected arrival time should always include a corrected speed (see MO11). Because we assume radial propagation, the component of the velocity is always the radial component away from the Sun, and the others are zero. For self-similar expansion, the following relation must be valid because the speed needs to be proportional to distance along the front, so the shape does not change with time:
\begin{eqnarray}
 \frac{d_i}{R_{iSSE}}=\frac{V_{iSSE}}{V_{SSE}},
\end{eqnarray}
Rearranging and substituting $R_{iSSE}$ from Equation~\ref{eq:risse} shows that the front at an angle $\Delta$ to the ICME apex moves with the following speed ($V_{iSSE}< V_{SSE}$) away from the Sun:  
\begin{eqnarray}\label{eq:vcorr}
V_{iSSE} = V_{SSE} \frac{   \cos \Delta +
 \sqrt{\sin^2 \lambda - \sin^2 \Delta} }{ (1+\sin\lambda)}
\end{eqnarray}
Again, for $\lambda=90$\degreee, Equation~\ref{eq:vcorr} reduces to: 
\begin{eqnarray}
V_{iHM}=V_{HM} \cos({\Delta})
\end{eqnarray}
as stated by MO11 for the Harmonic Mean case.

\section{Plots for arrival time and speed corrections}

In this section we explore the significance of the previously derived formulas by plotting the corrections for arrival time and speed for spacecraft positions along the ICME flank, for different sets of parameters. In Figure~\ref{size} we introduce five SSE fronts with angular half widths of $\lambda=[10,25,40,60,90]$\degreee. These values correspond to a radius of curvature, which is equal to the radius $r$ of the circle as given by Equation~\ref{eq:curvature}, of $r=[0.086, 0.211, 0.321, 0.433, 0.5]$~AU for an apex distance of $R_{SSE}=1$~AU. We will also use  $V_{SSE}=[400, 600, 800, 1000]$~km~s$^{-1}$, covering typical ICME speeds observed \emph{in situ} at 1~AU  throughout a solar cycle \cite{ric10}.


\subsection{Arrival time}
\begin{figure}[ht]
\centering
    \includegraphics[width=1\textwidth,clip=]{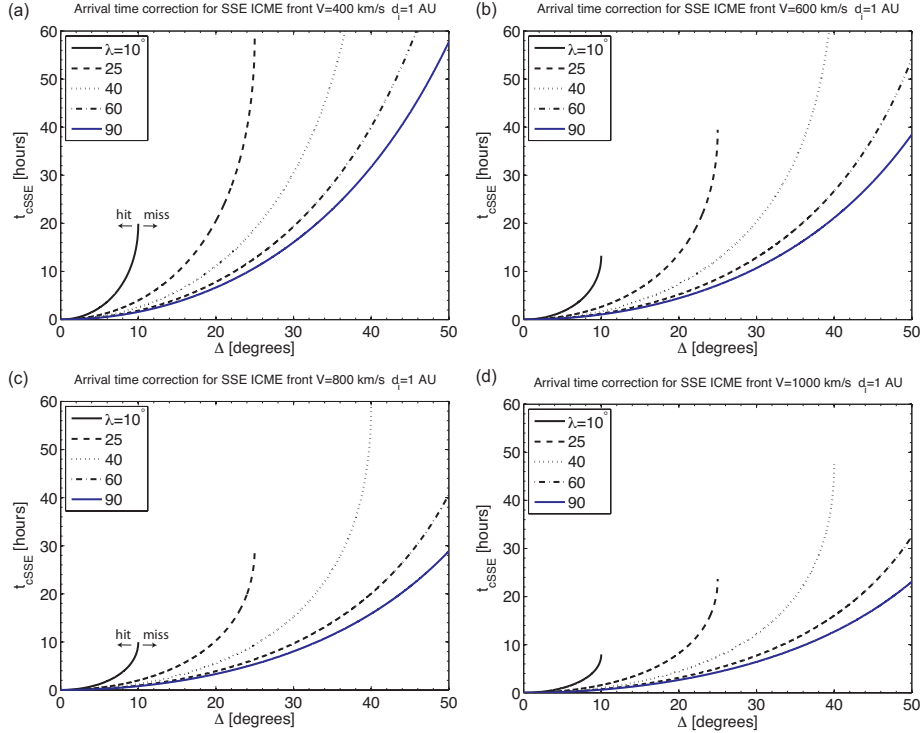}
    \caption{(a): The arrival time correction $t_{cSSE}$ assuming the ICME parameters $V_{SSE}=400$~\kmsec and the \emph{in situ} observer situated at $d_i=1$~AU, for different values of the ICME half width $\lambda$. (b) Similar to above, but for  $V_{SSE}=600$~\kmsec, (c) $V_{SSE}=800$~\kmsec, (d) $V_{SSE}=1000$~\kmsec. }\label{time}
\end{figure}

In Figure~\ref{time}a, we plot the arrival time correction $t_{cSSE}$ given by Equation~\ref{eq:tisse} as a function of $\Delta$ for an ICME speed of $V_{SSE}=400$~\kmsec ~and the \emph{in situ} spacecraft at Earth distance ($d_i=1$~AU). This is done similarly to MO11 (their Figure~8b), but these authors restricted themselves to the HM case ($\lambda=90$\degreee),  which is the blue solid curve in Figure~\ref{time}a.  The functions for $\lambda=[10,25,40,60]^{\circ}$ each stop at the same value for $\Delta$, though only seen in the plot for $\lambda$=10\degreee. This is  because for $\Delta> \lambda$ the ICME does not hit the spacecraft, and $t_{cSSE}$ becomes imaginary. For smaller values of $\lambda$, a few degrees difference in $\Delta$ can lead to large differences in the arrival time, whereas for large $\lambda$ values the curves becomes increasingly flatter.  Also note that for $\lambda=60^{\circ}$ there is already almost no difference to HM, so the formulas mainly affect cases where small values of $\lambda$ are combined with a $\Delta$ close to $\lambda$, which is - in other words - the case for glancing encounters of ICMEs which are of small angular extent along the line-of-sight. This is either the case for intrinsically narrow ICMEs or elongated flux-rope like ICMEs which have a high inclination of their axis to the ecliptic plane.
 
Figures~\ref{time}b, \ref{time}c, \ref{time}d show similar plots for $V_{SSE}=[600, 800, 1000]$~\kmsec, which allow to quickly look up, also useful for real-time assessments, how long the flank might be delayed for a given set of ICME parameters. As an example, it can be seen that for an average width of an ICME shock front (a full width of 80\degree corresponding to $\lambda=40$\degreee), the arrival time correction stays low ($< 10$ hours) for encounters within $\Delta< 25$\degreee,  but it quickly rises when $\Delta > 25$\degree and may easily delay the front arrival by one or even two days. Thus, the conclusion is that the arrival time for an ICME front approximated by SSE is increasingly hard to pinpoint at its flanks, where small errors in $\Delta$ may lead to large errors in the arrival time prediction.

\subsection{Arrival speed}

\begin{figure}[ht]
\centering
    \includegraphics[width=1\textwidth,clip=]{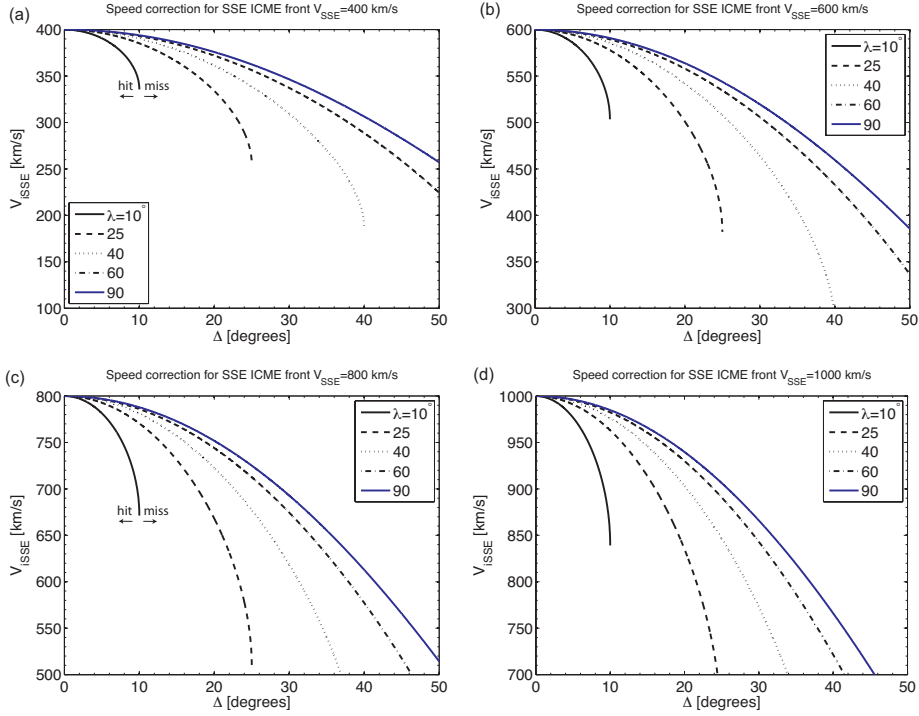}
    \caption{(a): The corrected ICME speed $V_{iSSE}$ at a longitudinal separation $\Delta$ between the ICME apex and the \emph{in situ} target, assuming the ICME parameters $V_{SSE}=400$~\kmsec  and the given angular half widths $\lambda$. (b) Similar to above, but for  $V_{SSE}=600$~\kmsec, (c) for $V_{SSE}=800$~\kmsec, (d) for $V_{SSE}=1000$~\kmsec.}\label{speed}
\end{figure}

For Figure~\ref{speed} we repeated this analysis for the speed corrections. We use the same sets of parameters as in the previous section to plug into Equation~\ref{eq:vcorr}. Note that $V_{iSSE}$ does not depend on $d_i,$ because the front always moves with constant speed, regardless of the heliospheric distance of $I$. Again, where $\Delta> \lambda$ the ICME does not hit the spacecraft. Clearly, the functional forms are governed by the same correction factor as compared to the arrival time in the previous section. Again, these plots are meant to be useful for looking up the effect of a glancing encounter for given parameters. For cases where $\Delta>30$\degreee, the arrival speed can be slower in the order of $100-200$ \kmsec~compared to the apex. Again, this shows that for geometrically consistent calculations of arrival times and speeds it is necessary to include these formulas because they lead to significant deviations from those inferred from the ICME apex.
 
\section{Conclusions}

Our aim was to show how arrival times and speeds can be analytically calculated for self-similar expanding circular fronts with constant angular width. We have shown plots of the corrections for speeds and arrival times for an observer at 1 AU and typical ICME speeds at 1 AU. This is not only useful for techniques which assume this geometry for application to observations by a single-spacecraft heliospheric imager (HI) instrument (SSEF, \citeauthor{dav12} \citeyear{dav12}), or two HIs \cite{lug10}, to forecast the effects of interplanetary coronal mass ejections at Earth and other planets, but for any empirical/analytical technique which is used to calculate arrival times of ICMEs (summary by \citeauthor{sis06}, \citeyear{sis06}; see also \citeauthor{vrs12} \citeyear{vrs12}, this issue, for the drag-based model). Assuming the SSE geometry leads to the following results: For glancing encounters of an ICME with a planet or spacecraft, the flank parts of an ICME might be delayed by one or - in extreme cases - two days, and speeds can be affected in the order of 100 to 200 \kmsec. These corrections are both significant, given that arrival time predictions are usually in the order of $\pm 12$ hours (e.g. \citeauthor{dav11} \citeyear{dav11}), and  geoeffects of ICMEs are thought to be determined by the  reconnection electric field $-V_r \times B_z$, depending on the plasma radial velocity component $V_r$ and the southward magnetic field component $B_z$ (e.g. \citeauthor{bur75}, \citeyear{bur75}). 

Further assumptions we used for the derived corrections are that after the front passed the heliocentric distance of the planet or spacecraft, (1) the ICME propagates radially away from the Sun and (2) it does so with constant speed. Clearly, in a structured solar wind, the ICME front might easily become distorted by the heliospheric current sheet, co-rotating interaction regions or other ICMEs, thus one always has to keep in mind that the geometry used here is a highly idealized one. We also note that in the SSE framework, the curvature of the front is coupled to the full width, leading to a rather rigid description of its shape. De-coupling the width from the curvature would be possible in an elliptical model (or a related geometry used by \citeauthor{sav11} \citeyear{sav11}), at the cost of introducing another free parameter. We leave these ideas for future studies. 

Nevertheless, the expressions presented in this paper allow to check the results of SSEF against ICME \emph{in situ} observations. It is indeed our aim to use the presented formulas in connection with the SSEF  technique and its limiting cases (FPF and HMF) to test the efficacy of using heliospheric imagers as a prediction tool for real time space weather forecasting by studying a larger number of events. To do so in a geometrically consistent way, the derived corrections will be applied to comparisons between ICME arrival times and speeds derived from STEREO/HI with SSEF and those observed \emph{in situ} by various spacecraft, extending other studies using elongation-fitting techniques and their stereoscopic versions (\citeauthor{dav09} \citeyear{dav09};  \citeauthor{liu10} \citeyear{liu10}a,b; \citeyear{liu11}; \citeauthor{moe09c} \citeyear{moe09c}, \citeyear{moe10}, \citeyear{moe11}; \citeauthor{lug12} \citeyear{lug12}).

%
 \begin{acks}
The presented work has received funding from the European Union Seventh
Framework Programme (FP7/2007-2013) under grant agreement n\degree 263252
[COMESEP]. This research was supported by a Marie Curie International Outgoing
Fellowship within the 7th European Community Framework Programme.
J.A.D acknowledges funding from the UK Space Agency. We also thank the guest editor and the referee for ideas on how to simplify the final equations and their derivation.
 \end{acks}

%
%
 \bibliographystyle{spr-mp-sola}

\begin{thebibliography}{44}
\ifx \bisbn   \undefined \def \bisbn  #1{ISBN #1}\fi
\ifx \binits  \undefined \def \binits#1{#1} \fi
\ifx \bauthor  \undefined \def \bauthor#1{#1} \fi
\ifx \batitle  \undefined \def \batitle#1{#1} \fi
\ifx \bjtitle  \undefined \def \bjtitle#1{\textit{#1}}\fi
\ifx \bvolume  \undefined \def \bvolume#1{\textbf{#1}}\fi
\ifx \byear  \undefined \def \byear#1{#1} \fi
\ifx \bissue  \undefined \def \bissue#1{#1} \fi
\ifx \bfpage  \undefined \def \bfpage#1{#1} \fi
\ifx \blpage  \undefined \def \blpage #1{#1} \fi
\ifx \burl  \undefined \def \burl#1{\textsf{#1}} \fi
\ifx \doiurl  \undefined \def \doiurl#1{\textsf{#1}} \fi
\ifx \betal  \undefined \def \betal{\textit{et al.}} \fi
\ifx \binstitute  \undefined \def \binstitute#1{#1} \fi
\ifx \bctitle  \undefined \def \bctitle#1{#1} \fi
\ifx \beditor  \undefined \def \beditor#1{#1} \fi
\ifx \bpublisher  \undefined \def \bpublisher#1{#1} \fi
\ifx \bbtitle  \undefined \def \bbtitle#1{\textit{#1}} \fi
\ifx \bedition  \undefined \def \bedition#1{#1} \fi
\ifx \bseriesno  \undefined \def \bseriesno#1{\textbf{#1}} \fi
\ifx \blocation  \undefined \def \blocation#1{#1} \fi
\ifx \bsertitle  \undefined \def \bsertitle#1{\textit{#1}} \fi
\ifx \bsnm \undefined \def \bsnm#1{#1} \fi
\ifx \bsuffix \undefined \def \bsuffix#1{#1} \fi
\ifx \bparticle \undefined \def \bparticle#1{#1} \fi
\ifx \barticle \undefined \def \barticle#1{#1} \fi
\ifx \botherref \undefined \def \botherref #1{#1} \fi
\ifx \url \undefined \def \url#1{\textsf{#1}} \fi
\ifx \bchapter \undefined \def \bchapter#1{#1} \fi
\ifx \bbook \undefined \def \bbook#1{#1} \fi
\ifx \bcomment \undefined \def \bcomment#1{#1} \fi
\ifx \oauthor \undefined \def \oauthor#1{#1} \fi
\ifx \citeauthoryear \undefined \def \citeauthoryear#1{#1} \fi
\def \endbibitem {}

\bibitem[\protect\citeauthoryear{{Burton}, {McPherron}, and
  {Russell}}{1975}]{bur75}
\begin{barticle}
\bauthor{\bsnm{{Burton}}, \binits{R.K.}}, \bauthor{\bsnm{{McPherron}},
  \binits{R.L.}}, \bauthor{\bsnm{{Russell}}, \binits{C.T.}}:
\byear{1975},
\batitle{{An empirical relationship between interplanetary conditions and
  Dst}}.
\bjtitle{\jgr}
\bvolume{80},
\bfpage{4204}.
doi:\doiurl{10.1029/JA080i031p04204}.
\end{barticle}
\endbibitem

\bibitem[\protect\citeauthoryear{{Byrne} \textit{et~al.}}{2010}]{byr10}
\begin{barticle}
\bauthor{\bsnm{{Byrne}}, \binits{J.P.}}, \bauthor{\bsnm{{Maloney}},
  \binits{S.A.}}, \bauthor{\bsnm{{McAteer}}, \binits{R.T.J.}},
  \bauthor{\bsnm{{Refojo}}, \binits{J.M.}}, \bauthor{\bsnm{{Gallagher}},
  \binits{P.T.}}:
\byear{2010},
\batitle{{Propagation of an Earth-directed coronal mass ejection in three
  dimensions}}.
\bjtitle{Nature Communications}
\bvolume{1}.
\bfpage{74}.
doi:\doiurl{10.1038/ncomms1077}.
\end{barticle}
\endbibitem

\bibitem[\protect\citeauthoryear{{Davies} \textit{et~al.}}{2009}]{dav09b}
\begin{barticle}
\bauthor{\bsnm{{Davies}}, \binits{J.A.}}, \bauthor{\bsnm{{Harrison}},
  \binits{R.A.}}, \bauthor{\bsnm{{Rouillard}}, \binits{A.P.}},
  \bauthor{\bsnm{{Sheeley}}, \binits{N.R.}}, \bauthor{\bsnm{{Perry}},
  \binits{C.H.}}, \bauthor{\bsnm{{Bewsher}}, \binits{D.}},
  \bauthor{\bsnm{{Davis}}, \binits{C.J.}}, \bauthor{\bsnm{{Eyles}},
  \binits{C.J.}}, \bauthor{\bsnm{{Crothers}}, \binits{S.R.}},
  \bauthor{\bsnm{{Brown}}, \binits{D.S.}}:
\byear{2009},
\batitle{{A synoptic view of solar transient evolution in the inner heliosphere
  using the Heliospheric Imagers on STEREO}}.
\bjtitle{\grl}
\bvolume{36},
\bfpage{2102}.
doi:\doiurl{10.1029/2008GL036182}.
\end{barticle}
\endbibitem


\bibitem[\protect\citeauthoryear{{Davies} \textit{et~al.}}{2012}]{dav12}
\begin{barticle}
\bauthor{\bsnm{{Davies}}, \binits{J.A.}}, \bauthor{\bsnm{{Harrison}},
  \binits{R.A.}}, \bauthor{\bsnm{{Perry}}, \binits{C.H.}},
  \bauthor{\bsnm{{M\"ostl}}, \binits{C.}}, \bauthor{\bsnm{{Lugaz}},
  \binits{N.}}, \bauthor{\bsnm{{Rollett}}, \binits{T.}},
  \bauthor{\bsnm{{Davis}}, \binits{C.J.}}, \bauthor{\bsnm{{Crothers}},
  \binits{S.}}, \bauthor{\bsnm{{Temmer}}, \binits{M.}},
  \bauthor{\bsnm{{Eyles}}, \binits{C.J.}}, \bauthor{\bsnm{{Savani}}, \binits{N.P.}}:
\byear{2012},
\batitle{{A self-similar expansion model for use in solar wind transient propagation studies}}.
\bjtitle{\apj}
\bvolume{},
\bfpage{in press}.
\end{barticle}
\endbibitem


\bibitem[\protect\citeauthoryear{{Davis} \textit{et~al.}}{2009}]{dav09}
\begin{barticle}
\bauthor{\bsnm{{Davis}}, \binits{C.J.}}, \bauthor{\bsnm{{Davies}},
  \binits{J.A.}}, \bauthor{\bsnm{{Lockwood}}, \binits{M.}},
  \bauthor{\bsnm{{Rouillard}}, \binits{A.P.}}, \bauthor{\bsnm{{Eyles}},
  \binits{C.J.}}, \bauthor{\bsnm{{Harrison}}, \binits{R.A.}}:
\byear{2009},
\batitle{{Stereoscopic imaging of an Earth-impacting solar coronal mass
  ejection: A major milestone for the STEREO mission}}.
\bjtitle{\grl}
\bvolume{36},
\bfpage{8102}.
doi:\doiurl{10.1029/2009GL038021}.
\end{barticle}
\endbibitem

\bibitem[\protect\citeauthoryear{{Davis} \textit{et~al.}}{2011}]{dav11}
\begin{barticle}
\bauthor{\bsnm{{Davis}}, \binits{C.J.}}, \bauthor{\bsnm{{de Koning}},
  \binits{C.A.}}, \bauthor{\bsnm{{Davies}}, \binits{J.A.}},
  \bauthor{\bsnm{{Biesecker}}, \binits{D.}}, \bauthor{\bsnm{{Millward}},
  \binits{G.}}, \bauthor{\bsnm{{Dryer}}, \binits{M.}}, \bauthor{\bsnm{{Deehr}},
  \binits{C.}}, \bauthor{\bsnm{{Webb}}, \binits{D.F.}},
  \bauthor{\bsnm{{Schenk}}, \binits{K.}}, \bauthor{\bsnm{{Freeland}},
  \binits{S.L.}}, \bauthor{\bsnm{{M{\"o}stl}}, \binits{C.}},
  \bauthor{\bsnm{{Farrugia}}, \binits{C.J.}}, \bauthor{\bsnm{{Odstrcil}},
  \binits{D.}}:
\byear{2011},
\batitle{{A comparison of space weather analysis techniques used to predict the
  arrival of the Earth-directed CME and its shockwave launched on 8 April
  2010}}.
\bjtitle{Space Weather}
\bvolume{9},
\bfpage{1005}.
doi:\doiurl{10.1029/2010SW000620}.
\end{barticle}
\endbibitem

\bibitem[\protect\citeauthoryear{{de Koning}, {Pizzo}, and
  {Biesecker}}{2009}]{dek09}
\begin{barticle}
\bauthor{\bsnm{{de Koning}}, \binits{C.A.}}, \bauthor{\bsnm{{Pizzo}},
  \binits{V.J.}}, \bauthor{\bsnm{{Biesecker}}, \binits{D.A.}}:
\byear{2009},
\batitle{{Geometric Localization of CMEs in 3D Space Using STEREO Beacon Data:
  First Results}}.
\bjtitle{\solphys}
\bvolume{256},
\bfpage{167}.
doi:\doiurl{10.1007/s11207-009-9344-7}.
\end{barticle}
\endbibitem

\bibitem[\protect\citeauthoryear{{Eyles} \textit{et~al.}}{2003}]{eyl03}
\begin{barticle}
\bauthor{\bsnm{{Eyles}}, \binits{C.J.}}, \bauthor{\bsnm{{Simnett}},
  \binits{G.M.}}, \bauthor{\bsnm{{Cooke}}, \binits{M.P.}},
  \bauthor{\bsnm{{Jackson}}, \binits{B.V.}}, \bauthor{\bsnm{{Buffington}},
  \binits{A.}}, \bauthor{\bsnm{{Hick}}, \binits{P.P.}},
  \bauthor{\bsnm{{Waltham}}, \binits{N.R.}}, \bauthor{\bsnm{{King}},
  \binits{J.M.}}, \bauthor{\bsnm{{Anderson}}, \binits{P.A.}},
  \bauthor{\bsnm{{Holladay}}, \binits{P.E.}}:
\byear{2003},
\batitle{{The Solar Mass Ejection Imager (Smei)}}.
\bjtitle{\solphys}
\bvolume{217},
\bfpage{319}.
\end{barticle}
\endbibitem

\bibitem[\protect\citeauthoryear{{Eyles} \textit{et~al.}}{2009}]{eyl09}
\begin{barticle}
\bauthor{\bsnm{{Eyles}}, \binits{C.J.}}, \bauthor{\bsnm{{Harrison}},
  \binits{R.A.}}, \bauthor{\bsnm{{Davis}}, \binits{C.J.}},
  \bauthor{\bsnm{{Waltham}}, \binits{N.R.}}, \bauthor{\bsnm{{Shaughnessy}},
  \binits{B.M.}}, \bauthor{\bsnm{{Mapson-Menard}}, \binits{H.C.A.}},
  \bauthor{\bsnm{{Bewsher}}, \binits{D.}}, \bauthor{\bsnm{{Crothers}},
  \binits{S.R.}}, \bauthor{\bsnm{{Davies}}, \binits{J.A.}},
  \bauthor{\bsnm{{Simnett}}, \binits{G.M.}}, \bauthor{\bsnm{{Howard}},
  \binits{R.A.}}, \bauthor{\bsnm{{Moses}}, \binits{J.D.}},
  \bauthor{\bsnm{{Newmark}}, \binits{J.S.}}, \bauthor{\bsnm{{Socker}},
  \binits{D.G.}}, \bauthor{\bsnm{{Halain}}, \binits{J.P.}},
  \bauthor{\bsnm{{Defise}}, \binits{J.M.}}, \bauthor{\bsnm{{Mazy}},
  \binits{E.}}, \bauthor{\bsnm{{Rochus}}, \binits{P.}}:
\byear{2009},
\batitle{{The Heliospheric Imagers Onboard the STEREO Mission}}.
\bjtitle{\solphys}
\bvolume{254},
\bfpage{387}.
doi:\doiurl{10.1007/s11207-008-9299-0}.
\end{barticle}
\endbibitem

\bibitem[\protect\citeauthoryear{{Gopalswamy} \textit{et~al.}}{2001}]{gop01}
\begin{barticle}
\bauthor{\bsnm{{Gopalswamy}}, \binits{N.}}, \bauthor{\bsnm{{Lara}},
  \binits{A.}}, \bauthor{\bsnm{{Yashiro}}, \binits{S.}},
  \bauthor{\bsnm{{Kaiser}}, \binits{M.L.}}, \bauthor{\bsnm{{Howard}},
  \binits{R.A.}}:
\byear{2001},
\batitle{{Predicting the 1-AU arrival times of coronal mass ejections}}.
\bjtitle{\jgr}
\bvolume{106},
\bfpage{29207}.
doi:\doiurl{10.1029/2001JA000177}.
\end{barticle}
\endbibitem

\bibitem[\protect\citeauthoryear{{Gopalswamy} \textit{et~al.}}{2011}]{gop11}
\begin{barticle}
\bauthor{\bsnm{{Gopalswamy}}, \binits{N.}}, \bauthor{\bsnm{{Davila}},
  \binits{J.M.}}, \bauthor{\bsnm{{St.~Cyr}}, \binits{O.C.}},
  \bauthor{\bsnm{{Sittler}}, \binits{E.C.}}, \bauthor{\bsnm{{Auch{\`e}re}},
  \binits{F.}}, \bauthor{\bsnm{{Duvall}}, \binits{T.L.}},
  \bauthor{\bsnm{{Hoeksema}}, \binits{J.T.}}, \bauthor{\bsnm{{Maksimovic}},
  \binits{M.}}, \bauthor{\bsnm{{MacDowall}}, \binits{R.J.}},
  \bauthor{\bsnm{{Szabo}}, \binits{A.}}, \bauthor{\bsnm{{Collier}},
  \binits{M.R.}}:
\byear{2011},
\batitle{{Earth-Affecting Solar Causes Observatory (EASCO): A potential
  International Living with a Star Mission from Sun-Earth L5}}.
\bjtitle{Journal of Atmospheric and Solar-Terrestrial Physics}
\bvolume{73},
\bfpage{658}.
doi:\doiurl{10.1016/j.jastp.2011.01.013}.
\end{barticle}
\endbibitem

\bibitem[\protect\citeauthoryear{{Howard} and {Tappin}}{2009}]{how09a}
\begin{barticle}
\bauthor{\bsnm{{Howard}}, \binits{T.A.}}, \bauthor{\bsnm{{Tappin}},
  \binits{S.J.}}:
\byear{2009},
\batitle{{Interplanetary Coronal Mass Ejections Observed in the Heliosphere: 1.
  Review of Theory}}.
\bjtitle{\ssr}
\bvolume{147},
\bfpage{31}.
doi:\doiurl{10.1007/s11214-009-9542-5}.
\end{barticle}
\endbibitem

\bibitem[\protect\citeauthoryear{{Kahler} and {Webb}}{2007}]{kah07}
\begin{barticle}
\bauthor{\bsnm{{Kahler}}, \binits{S.W.}}, \bauthor{\bsnm{{Webb}},
  \binits{D.F.}}:
\byear{2007},
\batitle{{V arc interplanetary coronal mass ejections observed with the Solar
  Mass Ejection Imager}}.
\bjtitle{Journal of Geophysical Research (Space Physics)}
\bvolume{112},
\bfpage{9103}.
doi:\doiurl{10.1029/2007JA012358}.
\end{barticle}
\endbibitem

\bibitem[\protect\citeauthoryear{{Kaiser} \textit{et~al.}}{2008}]{kai08}
\begin{barticle}
\bauthor{\bsnm{{Kaiser}}, \binits{M.L.}}, \bauthor{\bsnm{{Kucera}},
  \binits{T.A.}}, \bauthor{\bsnm{{Davila}}, \binits{J.M.}},
  \bauthor{\bsnm{{St.~Cyr}}, \binits{O.C.}}, \bauthor{\bsnm{{Guhathakurta}},
  \binits{M.}}, \bauthor{\bsnm{{Christian}}, \binits{E.}}:
\byear{2008},
\batitle{{The STEREO Mission: An Introduction}}.
\bjtitle{Space Science Reviews}
\bvolume{136},
\bfpage{5}.
doi:\doiurl{10.1007/s11214-007-9277-0}.
\end{barticle}
\endbibitem

\bibitem[\protect\citeauthoryear{{Kilpua} \textit{et~al.}}{2011}]{kil11}
\begin{barticle}
\bauthor{\bsnm{{Kilpua}}, \binits{E.K.J.}}, \bauthor{\bsnm{{Jian}},
  \binits{L.K.}}, \bauthor{\bsnm{{Li}}, \binits{Y.}},
  \bauthor{\bsnm{{Luhmann}}, \binits{J.G.}}, \bauthor{\bsnm{{Russell}},
  \binits{C.T.}}:
\byear{2011},
\batitle{{Multipoint ICME encounters: Pre-STEREO and STEREO observations}}.
\bjtitle{Journal of Atmospheric and Solar-Terrestrial Physics}
\bvolume{73},
\bfpage{1228}.
doi:\doiurl{10.1016/j.jastp.2010.10.012}.
\end{barticle}
\endbibitem

\bibitem[\protect\citeauthoryear{{Liu} \textit{et~al.}}{2010}]{liu10}
\begin{barticle}
\bauthor{\bsnm{{Liu}}, \binits{Y.}}, \bauthor{\bsnm{{Davies}}, \binits{J.A.}},
  \bauthor{\bsnm{{Luhmann}}, \binits{J.G.}}, \bauthor{\bsnm{{Vourlidas}},
  \binits{A.}}, \bauthor{\bsnm{{Bale}}, \binits{S.D.}}, \bauthor{\bsnm{{Lin}},
  \binits{R.P.}}:
\byear{2010a},
\batitle{{Geometric Triangulation of Imaging Observations to Track Coronal Mass
  Ejections Continuously Out to 1 AU}}.
\bjtitle{\apjl}
\bvolume{710},
\bfpage{L82}.
doi:\doiurl{10.1088/2041-8205/710/1/L82}.
\end{barticle}
\endbibitem

\bibitem[\protect\citeauthoryear{{Liu} \textit{et~al.}}{2010}]{liu10b}
\begin{barticle}
\bauthor{\bsnm{{Liu}}, \binits{Y.}}, \bauthor{\bsnm{{Thernisien}},
  \binits{A.}}, \bauthor{\bsnm{{Luhmann}}, \binits{J.G.}},
  \bauthor{\bsnm{{Vourlidas}}, \binits{A.}}, \bauthor{\bsnm{{Davies}},
  \binits{J.A.}}, \bauthor{\bsnm{{Lin}}, \binits{R.P.}},
  \bauthor{\bsnm{{Bale}}, \binits{S.D.}}:
\byear{2010b},
\batitle{{Reconstructing Coronal Mass Ejections with Coordinated Imaging and in
  Situ Observations: Global Structure, Kinematics, and Implications for Space
  Weather Forecasting}}.
\bjtitle{\apj}
\bvolume{722},
\bfpage{1762}.
doi:\doiurl{10.1088/0004-637X/722/2/1762}.
\end{barticle}
\endbibitem

\bibitem[\protect\citeauthoryear{{Liu} \textit{et~al.}}{2011}]{liu11}
\begin{barticle}
\bauthor{\bsnm{{Liu}}, \binits{Y.}}, \bauthor{\bsnm{{Luhmann}}, \binits{J.G.}},
  \bauthor{\bsnm{{Bale}}, \binits{S.D.}}, \bauthor{\bsnm{{Lin}},
  \binits{R.P.}}:
\byear{2011},
\batitle{{Solar Source and Heliospheric Consequences of the 2010 April 3
  Coronal Mass Ejection: A Comprehensive View}}.
\bjtitle{\apj}
\bvolume{734},
\bfpage{84}.
doi:\doiurl{10.1088/0004-637X/734/2/84}.
\end{barticle}
\endbibitem

\bibitem[\protect\citeauthoryear{{Lugaz}}{2010}]{lug10b}
\begin{barticle}
\bauthor{\bsnm{{Lugaz}}, \binits{N.}}:
\byear{2010},
\batitle{{Accuracy and Limitations of Fitting and Stereoscopic Methods to
  Determine the Direction of Coronal Mass Ejections from Heliospheric Imagers
  Observations}}.
\bjtitle{\solphys}
\bvolume{267},
\bfpage{411}.
doi:\doiurl{10.1007/s11207-010-9654-9}.
\end{barticle}
\endbibitem


\bibitem[\protect\citeauthoryear{{Lugaz} \textit{et~al.}}{2010}]{lug10}
\begin{barticle}
\bauthor{\bsnm{{Lugaz}}, \binits{N.}}, \bauthor{\bsnm{{Hernandez-Charpak}},
  \binits{J.N.}}, \bauthor{\bsnm{{Roussev}}, \binits{I.I.}},
  \bauthor{\bsnm{{Davis}}, \binits{C.J.}}, \bauthor{\bsnm{{Vourlidas}},
  \binits{A.}}, \bauthor{\bsnm{{Davies}}, \binits{J.A.}}:
\byear{2010},
\batitle{{Determining the Azimuthal Properties of Coronal Mass Ejections from
  Multi-Spacecraft Remote-Sensing Observations with STEREO SECCHI}}.
\bjtitle{\apj}
\bvolume{715},
\bfpage{493}.
doi:\doiurl{10.1088/0004-637X/715/1/493}.
\end{barticle}
\endbibitem



\bibitem[\protect\citeauthoryear{{Lugaz} \textit{et~al.}}{2012}]{lug12}
\begin{barticle}
\bauthor{\bsnm{{Lugaz}}, \binits{N.}},  \bauthor{\bsnm{{Kintner}}, \binits{P.}}, \bauthor{\bsnm{{M\"ostl}}, \binits{C.}}, \bauthor{\bsnm{{Jian}}, \binits{L.K.}}, \bauthor{\bsnm{{Davis}}, \binits{C.J.}}, \bauthor{\bsnm{{Farrugia}}, \binits{C.J.}}:
\byear{2012},
\batitle{{Heliospheric Observations of STEREO-Directed Coronal Mass Ejections in 2008-2010: Lessons for Future Observations of Earth-Directed CMEs}}.
\bjtitle{Sol. Phys.}
\bvolume{},
\bfpage{submitted}.
\end{barticle}
\endbibitem



\bibitem[\protect\citeauthoryear{{Lugaz}, {Roussev}, and
  {Gombosi}}{2011}]{lug11}
\begin{barticle}
\bauthor{\bsnm{{Lugaz}}, \binits{N.}}, \bauthor{\bsnm{{Roussev}},
  \binits{I.I.}}, \bauthor{\bsnm{{Gombosi}}, \binits{T.I.}}:
\byear{2011},
\batitle{{Determining CME parameters by fitting heliospheric observations:
  Numerical investigation of the accuracy of the methods}}.
\bjtitle{Advances in Space Research}
\bvolume{48},
\bfpage{292}.
doi:\doiurl{10.1016/j.asr.2011.03.015}.
\end{barticle}
\endbibitem

\bibitem[\protect\citeauthoryear{{Lugaz}, {Vourlidas}, and
  {Roussev}}{2009}]{lug09a}
\begin{barticle}
\bauthor{\bsnm{{Lugaz}}, \binits{N.}}, \bauthor{\bsnm{{Vourlidas}},
  \binits{A.}}, \bauthor{\bsnm{{Roussev}}, \binits{I.I.}}:
\byear{2009},
\batitle{{Deriving the radial distances of wide coronal mass ejections from
  elongation measurements in the heliosphere - application to CME-CME
  interaction}}.
\bjtitle{Annales Geophysicae}
\bvolume{27},
\bfpage{3479}.
\end{barticle}
\endbibitem


\bibitem[\protect\citeauthoryear{{Maloney} and {Gallagher}}{2010}]{mal10}
\begin{barticle}
\bauthor{\bsnm{{Maloney}}, \binits{S.A.}}, \bauthor{\bsnm{{Gallagher}},
  \binits{P.T.}}:
\byear{2010},
\batitle{{Solar Wind Drag and the Kinematics of Interplanetary Coronal Mass
  Ejections}}.
\bjtitle{\apjl}
\bvolume{724},
\bfpage{L127}.
doi:\doiurl{10.1088/2041-8205/724/2/L127}.
\end{barticle}
\endbibitem

\bibitem[\protect\citeauthoryear{{Mierla} \textit{et~al.}}{2010}]{mie10}
\begin{barticle}
\bauthor{\bsnm{{Mierla}}, \binits{M.}}, \bauthor{\bsnm{{Inhester}},
  \binits{B.}}, \bauthor{\bsnm{{Antunes}}, \binits{A.}},
  \bauthor{\bsnm{{Boursier}}, \binits{Y.}}, \bauthor{\bsnm{{Byrne}},
  \binits{J.P.}}, \bauthor{\bsnm{{Colaninno}}, \binits{R.}},
  \bauthor{\bsnm{{Davila}}, \binits{J.}}, \bauthor{\bsnm{{de Koning}},
  \binits{C.A.}}, \bauthor{\bsnm{{Gallagher}}, \binits{P.T.}},
  \bauthor{\bsnm{{Gissot}}, \binits{S.}}, \bauthor{\bsnm{{Howard}},
  \binits{R.A.}}, \bauthor{\bsnm{{Howard}}, \binits{T.A.}},
  \bauthor{\bsnm{{Kramar}}, \binits{M.}}, \bauthor{\bsnm{{Lamy}}, \binits{P.}},
  \bauthor{\bsnm{{Liewer}}, \binits{P.C.}}, \bauthor{\bsnm{{Maloney}},
  \binits{S.}}, \bauthor{\bsnm{{Marqu{\'e}}}, \binits{C.}},
  \bauthor{\bsnm{{McAteer}}, \binits{R.T.J.}}, \bauthor{\bsnm{{Moran}},
  \binits{T.}}, \bauthor{\bsnm{{Rodriguez}}, \binits{L.}},
  \bauthor{\bsnm{{Srivastava}}, \binits{N.}}, \bauthor{\bsnm{{St.~Cyr}},
  \binits{O.C.}}, \bauthor{\bsnm{{Stenborg}}, \binits{G.}},
  \bauthor{\bsnm{{Temmer}}, \binits{M.}}, \bauthor{\bsnm{{Thernisien}},
  \binits{A.}}, \bauthor{\bsnm{{Vourlidas}}, \binits{A.}},
  \bauthor{\bsnm{{West}}, \binits{M.J.}}, \bauthor{\bsnm{{Wood}},
  \binits{B.E.}}, \bauthor{\bsnm{{Zhukov}}, \binits{A.N.}}:
\byear{2010},
\batitle{{On the 3-D reconstruction of Coronal Mass Ejections using coronagraph
  data}}.
\bjtitle{Annales Geophysicae}
\bvolume{28},
\bfpage{203}.
doi:\doiurl{10.5194/angeo-28-203-2010}.
\end{barticle}
\endbibitem

\bibitem[\protect\citeauthoryear{{M{\"o}stl} \textit{et~al.}}{2009}]{moe09c}
\begin{barticle}
\bauthor{\bsnm{{M{\"o}stl}}, \binits{C.}}, \bauthor{\bsnm{{Farrugia}},
  \binits{C.J.}}, \bauthor{\bsnm{{Temmer}}, \binits{M.}},
  \bauthor{\bsnm{{Miklenic}}, \binits{C.}}, \bauthor{\bsnm{{Veronig}},
  \binits{A.M.}}, \bauthor{\bsnm{{Galvin}}, \binits{A.B.}},
  \bauthor{\bsnm{{Leitner}}, \binits{M.}}, \bauthor{\bsnm{{Biernat}},
  \binits{H.K.}}:
\byear{2009},
\batitle{{Linking Remote Imagery of a Coronal Mass Ejection to Its \emph{in situ}
  Signatures at 1 AU}}.
\bjtitle{\apjl}
\bvolume{705},
\bfpage{L180}.
doi:\doiurl{10.1088/0004-637X/705/2/L180}.
\end{barticle}
\endbibitem


\bibitem[\protect\citeauthoryear{{M{\"o}stl} \textit{et~al.}}{2011}]{moe11}
\begin{barticle}
\bauthor{\bsnm{{M{\"o}stl}}, \binits{C.}}, \bauthor{\bsnm{{Rollett}},
  \binits{T.}}, \bauthor{\bsnm{{Lugaz}}, \binits{N.}},
  \bauthor{\bsnm{{Farrugia}}, \binits{C.J.}}, \bauthor{\bsnm{{Davies}},
  \binits{J.A.}}, \bauthor{\bsnm{{Temmer}}, \binits{M.}},
  \bauthor{\bsnm{{Veronig}}, \binits{A.M.}}, \bauthor{\bsnm{{Harrison}},
  \binits{R.A.}}, \bauthor{\bsnm{{Crothers}}, \binits{S.}},
  \bauthor{\bsnm{{Luhmann}}, \binits{J.G.}}, \bauthor{\bsnm{{Galvin}},
  \binits{A.B.}}, \bauthor{\bsnm{{Zhang}}, \binits{T.L.}},
  \bauthor{\bsnm{{Baumjohann}}, \binits{W.}}, \bauthor{\bsnm{{Biernat}},
  \binits{H.K.}}:
\byear{2011},
\batitle{{Arrival Time Calculation for Interplanetary Coronal Mass Ejections
  with Circular Fronts and Application to STEREO Observations of the 2009
  February 13 Eruption}}.
\bjtitle{\apj}
\bvolume{741},
\bfpage{34}.
doi:\doiurl{10.1088/0004-637X/741/1/34}.
\end{barticle}
\endbibitem

\bibitem[\protect\citeauthoryear{{M{\"o}stl} \textit{et~al.}}{2010}]{moe10}
\begin{barticle}
\bauthor{\bsnm{{M{\"o}stl}}, \binits{C.}}, \bauthor{\bsnm{{Temmer}},
  \binits{M.}}, \bauthor{\bsnm{{Rollett}}, \binits{T.}},
  \bauthor{\bsnm{{Farrugia}}, \binits{C.J.}}, \bauthor{\bsnm{{Liu}},
  \binits{Y.}}, \bauthor{\bsnm{{Veronig}}, \binits{A.M.}},
  \bauthor{\bsnm{{Leitner}}, \binits{M.}}, \bauthor{\bsnm{{Galvin}},
  \binits{A.B.}}, \bauthor{\bsnm{{Biernat}}, \binits{H.K.}}:
\byear{2010},
\batitle{{STEREO and Wind observations of a fast ICME flank triggering a
  prolonged geomagnetic storm on 5-7 April 2010}}.
\bjtitle{\grl}
\bvolume{37},
\bfpage{L24103}.
doi:\doiurl{10.1029/2010GL045175}.
\end{barticle}
\endbibitem

\bibitem[\protect\citeauthoryear{{Richardson} and {Cane}}{1993}]{ric93}
\begin{barticle}
\bauthor{\bsnm{{Richardson}}, \binits{I.G.}}, \bauthor{\bsnm{{Cane}},
  \binits{H.V.}}:
\byear{1993},
\batitle{{Signatures of shock drivers in the solar wind and their dependence on
  the solar source location}}.
\bjtitle{\jgr}
\bvolume{98},
\bfpage{15295}.
\end{barticle}
\endbibitem

\bibitem[\protect\citeauthoryear{{Richardson} and {Cane}}{2010}]{ric10}
\begin{barticle}
\bauthor{\bsnm{{Richardson}}, \binits{I.G.}}, \bauthor{\bsnm{{Cane}},
  \binits{H.V.}}:
\byear{2010},
\batitle{{Near-Earth Interplanetary Coronal Mass Ejections During Solar Cycle
  23 (1996--2009): Catalog and Summary of Properties}}.
\bjtitle{\solphys}
\bvolume{264},
\bfpage{189}.
doi:\doiurl{10.1007/s11207-010-9568-6}.
\end{barticle}
\endbibitem

\bibitem[\protect\citeauthoryear{{Rodriguez} \textit{et~al.}}{2011}]{rod11}
\begin{barticle}
\bauthor{\bsnm{{Rodriguez}}, \binits{L.}}, \bauthor{\bsnm{{Mierla}},
  \binits{M.}}, \bauthor{\bsnm{{Zhukov}}, \binits{A.N.}},
  \bauthor{\bsnm{{West}}, \binits{M.}}, \bauthor{\bsnm{{Kilpua}}, \binits{E.}}:
\byear{2011},
\batitle{{Linking Remote-Sensing and \emph{in situ} Observations of Coronal Mass
  Ejections Using STEREO}}.
\bjtitle{\solphys}
\bvolume{270},
\bfpage{561}.
doi:\doiurl{10.1007/s11207-011-9784-8}.
\end{barticle}
\endbibitem

\bibitem[\protect\citeauthoryear{{Rouillard} \textit{et~al.}}{2008}]{rou08}
\begin{barticle}
\bauthor{\bsnm{{Rouillard}}, \binits{A.P.}}, \bauthor{\bsnm{{Davies}},
  \binits{J.A.}}, \bauthor{\bsnm{{Forsyth}}, \binits{R.J.}},
  \bauthor{\bsnm{{Rees}}, \binits{A.}}, \bauthor{\bsnm{{Davis}},
  \binits{C.J.}}, \bauthor{\bsnm{{Harrison}}, \binits{R.A.}},
  \bauthor{\bsnm{{Lockwood}}, \binits{M.}}, \bauthor{\bsnm{{Bewsher}},
  \binits{D.}}, \bauthor{\bsnm{{Crothers}}, \binits{S.R.}},
  \bauthor{\bsnm{{Eyles}}, \binits{C.J.}}, \bauthor{\bsnm{{Hapgood}},
  \binits{M.}}, \bauthor{\bsnm{{Perry}}, \binits{C.H.}}:
\byear{2008},
\batitle{{First imaging of corotating interaction regions using the STEREO
  spacecraft}}.
\bjtitle{\grl}
\bvolume{35},
\bfpage{10110}.
doi:\doiurl{10.1029/2008GL033767}.
\end{barticle}
\endbibitem

\bibitem[\protect\citeauthoryear{{Rouillard} \textit{et~al.}}{2009}]{rou09}
\begin{barticle}
\bauthor{\bsnm{{Rouillard}}, \binits{A.P.}}, \bauthor{\bsnm{{Davies}},
  \binits{J.A.}}, \bauthor{\bsnm{{Forsyth}}, \binits{R.J.}},
  \bauthor{\bsnm{{Savani}}, \binits{N.P.}}, \bauthor{\bsnm{{Sheeley}},
  \binits{N.R.}}, \bauthor{\bsnm{{Thernisien}}, \binits{A.}},
  \bauthor{\bsnm{{Zhang}}, \binits{T.}}, \bauthor{\bsnm{{Howard}},
  \binits{R.A.}}, \bauthor{\bsnm{{Anderson}}, \binits{B.}},
  \bauthor{\bsnm{{Carr}}, \binits{C.M.}}, \bauthor{\bsnm{{Tsang}},
  \binits{S.}}, \bauthor{\bsnm{{Lockwood}}, \binits{M.}},
  \bauthor{\bsnm{{Davis}}, \binits{C.J.}}, \bauthor{\bsnm{{Harrison}},
  \binits{R.A.}}, \bauthor{\bsnm{{Bewsher}}, \binits{D.}},
  \bauthor{\bsnm{{Fr{\"a}nz}}, \binits{M.}}, \bauthor{\bsnm{{Crothers}},
  \binits{S.R.}}, \bauthor{\bsnm{{Eyles}}, \binits{C.J.}},
  \bauthor{\bsnm{{Brown}}, \binits{D.S.}}, \bauthor{\bsnm{{Whittaker}},
  \binits{I.}}, \bauthor{\bsnm{{Hapgood}}, \binits{M.}},
  \bauthor{\bsnm{{Coates}}, \binits{A.J.}}, \bauthor{\bsnm{{Jones}},
  \binits{G.H.}}, \bauthor{\bsnm{{Grande}}, \binits{M.}},
  \bauthor{\bsnm{{Frahm}}, \binits{R.A.}}, \bauthor{\bsnm{{Winningham}},
  \binits{J.D.}}:
\byear{2009},
\batitle{{A solar storm observed from the Sun to Venus using the STEREO, Venus
  Express, and MESSENGER spacecraft}}.
\bjtitle{Journal of Geophysical Research (Space Physics)}
\bvolume{114},
\bfpage{7106}.
doi:\doiurl{10.1029/2008JA014034}.
\end{barticle}
\endbibitem

\bibitem[\protect\citeauthoryear{{Savani} \textit{et~al.}}{2010}]{sav10}
\begin{barticle}
\bauthor{\bsnm{{Savani}}, \binits{N.P.}}, \bauthor{\bsnm{{Owens}},
  \binits{M.J.}}, \bauthor{\bsnm{{Rouillard}}, \binits{A.P.}},
  \bauthor{\bsnm{{Forsyth}}, \binits{R.J.}}, \bauthor{\bsnm{{Davies}},
  \binits{J.A.}}:
\byear{2010},
\batitle{{Observational Evidence of a Coronal Mass Ejection Distortion Directly
  Attributable to a Structured Solar Wind}}.
\bjtitle{\apjl}
\bvolume{714},
\bfpage{L128}.
doi:\doiurl{10.1088/2041-8205/714/1/L128}.
\end{barticle}
\endbibitem

\bibitem[\protect\citeauthoryear{{Savani} \textit{et~al.}}{2011}]{sav11}
\begin{barticle}
\bauthor{\bsnm{{Savani}}, \binits{N.P.}}, \bauthor{\bsnm{{Owens}},
  \binits{M.J.}}, \bauthor{\bsnm{{Rouillard}}, \binits{A.P.}},
  \bauthor{\bsnm{{Forsyth}}, \binits{R.J.}}, \bauthor{\bsnm{{Kusano}},
  \binits{K.}}, \bauthor{\bsnm{{Shiota}}, \binits{D.}},
  \bauthor{\bsnm{{Kataoka}}, \binits{R.}}:
\byear{2011},
\batitle{{Evolution of Coronal Mass Ejection Morphology with Increasing
  Heliocentric Distance. I. Geometrical Analysis}}.
\bjtitle{\apj}
\bvolume{731},
\bfpage{109}.
doi:\doiurl{10.1088/0004-637X/731/2/109}.
\end{barticle}
\endbibitem


\bibitem[\protect\citeauthoryear{{Savani} \textit{et~al.}}{2009}]{sav09}
\begin{barticle}
\bauthor{\bsnm{{Savani}}, \binits{N.P.}}, \bauthor{\bsnm{{Rouillard}},
  \binits{A.P.}}, \bauthor{\bsnm{{Davies}}, \binits{J.A.}},
  \bauthor{\bsnm{{Owens}}, \binits{M.J.}}, \bauthor{\bsnm{{Forsyth}},
  \binits{R.J.}}, \bauthor{\bsnm{{Davis}}, \binits{C.J.}},
  \bauthor{\bsnm{{Harrison}}, \binits{R.A.}}:
\byear{2009},
\batitle{{The radial width of a Coronal Mass Ejection between 0.1 and 0.4 AU
  estimated from the Heliospheric Imager on STEREO}}.
\bjtitle{Annales Geophysicae}
\bvolume{27},
\bfpage{4349}.
doi:\doiurl{10.5194/angeo-27-4349-2009}.
\end{barticle}
\endbibitem


\bibitem[\protect\citeauthoryear{{Schwenn} \textit{et~al.}}{2005}]{sch05}
\begin{barticle}
\bauthor{\bsnm{{Schwenn}}, \binits{R.}}, \bauthor{\bsnm{{dal Lago}},
  \binits{A.}}, \bauthor{\bsnm{{Huttunen}}, \binits{E.}},
  \bauthor{\bsnm{{Gonzalez}}, \binits{W.D.}}:
\byear{2005},
\batitle{{The association of coronal mass ejections with their effects near the
  Earth}}.
\bjtitle{Annales Geophysicae}
\bvolume{23},
\bfpage{1033}.
\end{barticle}
\endbibitem

\bibitem[\protect\citeauthoryear{{Sheeley} \textit{et~al.}}{1999}]{she99}
\begin{barticle}
\bauthor{\bsnm{{Sheeley}}, \binits{N.R.}}, \bauthor{\bsnm{{Walters}},
  \binits{J.H.}}, \bauthor{\bsnm{{Wang}}, \binits{Y.}},
  \bauthor{\bsnm{{Howard}}, \binits{R.A.}}:
\byear{1999},
\batitle{{Continuous tracking of coronal outflows: Two kinds of coronal mass
  ejections}}.
\bjtitle{\jgr}
\bvolume{104},
\bfpage{24739}.
doi:\doiurl{10.1029/1999JA900308}.
\end{barticle}
\endbibitem

\bibitem[\protect\citeauthoryear{{Siscoe} and {Schwenn}}{2006}]{sis06}
\begin{barticle}
\bauthor{\bsnm{{Siscoe}}, \binits{G.}}, \bauthor{\bsnm{{Schwenn}},
  \binits{R.}}:
\byear{2006},
\batitle{{CME Disturbance Forecasting}}.
\bjtitle{\ssr}
\bvolume{123},
\bfpage{453}.
doi:\doiurl{10.1007/s11214-006-9024-y}.
\end{barticle}
\endbibitem

\bibitem[\protect\citeauthoryear{{Tappin} and {Howard}}{2009}]{tap09}
\begin{barticle}
\bauthor{\bsnm{{Tappin}}, \binits{S.J.}}, \bauthor{\bsnm{{Howard}},
  \binits{T.A.}}:
\byear{2009},
\batitle{{Interplanetary Coronal Mass Ejections Observed in the Heliosphere: 2.
  Model and Data Comparison}}.
\bjtitle{\ssr}
\bvolume{147},
\bfpage{55}.
doi:\doiurl{10.1007/s11214-009-9550-5}.
\end{barticle}
\endbibitem

\bibitem[\protect\citeauthoryear{{Temmer}, {Preiss}, and
  {Veronig}}{2009}]{tem09}
\begin{barticle}
\bauthor{\bsnm{{Temmer}}, \binits{M.}}, \bauthor{\bsnm{{Preiss}}, \binits{S.}},
  \bauthor{\bsnm{{Veronig}}, \binits{A.M.}}:
\byear{2009},
\batitle{{CME Projection Effects Studied with STEREO/COR and SOHO/LASCO}}.
\bjtitle{\solphys}
\bvolume{256},
\bfpage{183}.
doi:\doiurl{10.1007/s11207-009-9336-7}.
\end{barticle}
\endbibitem

\bibitem[\protect\citeauthoryear{{Thernisien}, {Vourlidas}, and
  {Howard}}{2009}]{the09}
\begin{barticle}
\bauthor{\bsnm{{Thernisien}}, \binits{A.}}, \bauthor{\bsnm{{Vourlidas}},
  \binits{A.}}, \bauthor{\bsnm{{Howard}}, \binits{R.A.}}:
\byear{2009},
\batitle{{Forward Modeling of Coronal Mass Ejections Using STEREO/SECCHI
  Data}}.
\bjtitle{\solphys}
\bvolume{256},
\bfpage{130}.
doi:\doiurl{10.1007/s11207-009-9346-5}.
\end{barticle}
\endbibitem

\bibitem[\protect\citeauthoryear{{Vr{\v s}nak} and {Gopalswamy}}{2002}]{vrs02}
\begin{barticle}
\bauthor{\bsnm{{Vr{\v s}nak}}, \binits{B.}}, \bauthor{\bsnm{{Gopalswamy}},
  \binits{N.}}:
\byear{2002},
\batitle{{Influence of the aerodynamic drag on the motion of interplanetary
  ejecta}}.
\bjtitle{Journal of Geophysical Research (Space Physics)}
\bvolume{107},
\bfpage{1019}.
doi:\doiurl{10.1029/2001JA000120}.
\end{barticle}
\endbibitem

\bibitem[\protect\citeauthoryear{{Vr{\v s}nak} \textit{et~al.}}{2010}]{vrs10}
\begin{barticle}
\bauthor{\bsnm{{Vr{\v s}nak}}, \binits{B.}}, \bauthor{\bsnm{{{\v Z}ic}},
  \binits{T.}}, \bauthor{\bsnm{{Falkenberg}}, \binits{T.V.}},
  \bauthor{\bsnm{{M{\"o}stl}}, \binits{C.}}, \bauthor{\bsnm{{Vennerstrom}},
  \binits{S.}}, \bauthor{\bsnm{{Vrbanec}}, \binits{D.}}:
\byear{2010},
\batitle{{The role of aerodynamic drag in propagation of interplanetary coronal
  mass ejections}}.
\bjtitle{\aap}
\bvolume{512},
\bfpage{A43}.
doi:\doiurl{10.1051/0004-6361/200913482}.
\end{barticle}
\endbibitem




\bibitem[\protect\citeauthoryear{{Vr{\v s}nak} \textit{et~al.}}{2012}]{vrs12}
\begin{barticle}
\bauthor{\bsnm{{Vr{\v s}nak}}, \binits{B.}}, \bauthor{\bsnm{{{\v Z}ic}},
  \binits{T.}}, \bauthor{\bsnm{{Vrbanec}}, \binits{D.}}, \bauthor{\bsnm{{Dumbovi\'c}}, \binits{D.}}, \bauthor{\bsnm{{Veronig}}, \binits{A.}}, \bauthor{\bsnm{{Temmer}}, \binits{M.}}, \bauthor{\bsnm{{M{\"o}stl}}, \binits{C.}}, \bauthor{\bsnm{{Rollett}},
  \binits{T.}}, \bauthor{\bsnm{{Moon}}, \binits{Y.-J.}}, \bauthor{\bsnm{{Luli\'c}}, \binits{S.}}, \bauthor{\bsnm{{Shanmugaraju}}, \binits{A.}}:
\byear{2012},
\batitle{{Propagation of Interplanetary Coronal Mass
Ejections: the Drag-Based Model}}.
\bjtitle{Solar Phys.}
\bvolume{},
\bfpage{submitted}.
\end{barticle}
\endbibitem

\bibitem[\protect\citeauthoryear{{Wood}, {Howard}, and {Socker}}{2010}]{woo10}
\begin{barticle}
\bauthor{\bsnm{{Wood}}, \binits{B.E.}}, \bauthor{\bsnm{{Howard}},
  \binits{R.A.}}, \bauthor{\bsnm{{Socker}}, \binits{D.G.}}:
\byear{2010},
\batitle{{Reconstructing the Morphology of an Evolving Coronal Mass Ejection}}.
\bjtitle{\apj}
\bvolume{715},
\bfpage{1524}.
doi:\doiurl{10.1088/0004-637X/715/2/1524}.
\end{barticle}
\endbibitem

\bibitem[\protect\citeauthoryear{{Yashiro} \textit{et~al.}}{2004}]{yas04}
\begin{barticle}
\bauthor{\bsnm{{Yashiro}}, \binits{S.}}, \bauthor{\bsnm{{Gopalswamy}},
  \binits{N.}}, \bauthor{\bsnm{{Michalek}}, \binits{G.}},
  \bauthor{\bsnm{{St.~Cyr}}, \binits{O.C.}}, \bauthor{\bsnm{{Plunkett}},
  \binits{S.P.}}, \bauthor{\bsnm{{Rich}}, \binits{N.B.}},
  \bauthor{\bsnm{{Howard}}, \binits{R.A.}}:
\byear{2004},
\batitle{{A catalog of white light coronal mass ejections observed by the SOHO
  spacecraft}}.
\bjtitle{Journal of Geophysical Research (Space Physics)}
\bvolume{109},
\bfpage{7105}.
doi:\doiurl{10.1029/2003JA010282}.
\end{barticle}
\endbibitem

\end{thebibliography}
%

%
%
%

\end{article}
\end{document}